\documentclass[dvips,, 12pt]{article}
\usepackage{latexsym, epsfig, amsmath, graphicx, amssymb}
\usepackage{mathrsfs, amsfonts, graphics}
\begin{document}

\begin{center}
\textbf{\Large Uniqueness of Flat Spherically Symmetric Spacelike
Hypersurfaces Admitted by Spherically Symmetric Static Spactimes}\\[1.2cm]
Robert Beig$^{1}$ and Azad A. Siddiqui$^{2}$\\[3ex]
$^{1}$Gravitational Physics, Faculty of Physics, University of
Vienna, A-1090 Vienna, Austria, E-mail: Robert.Beig@univie.ac.at
\\[0pt]
$^{2}$Department of Basic Science and Humanities, EME College, National University
of Science and Technology, Peshawar Road, Rawalpindi, Pakistan,\\[0pt]
E-mail: azad@ceme.edu.pk \\[0pt]

\bigskip

\bigskip

\textbf{Abstract}
\end{center}

\begin{quotation}
It is known that spherically symmetric static spacetimes admit a
foliation by flat hypersurfaces. Such foliations have explicitly
been constructed for some spacetimes, using different approaches,
but none of them have proved or even discussed the uniqueness of
these foliations. The issue of uniqueness becomes more important
due to suitability of flat foliations for studying black hole
physics. Here flat spherically symmetric spacelike hypersurfaces
are obtained by a direct method. It is found that spherically
symmetric static spacetimes admit flat spherically symmetric
hypersurfaces, and that these hypersurfaces are unique up to
translation under the timelike Killing vector. This result
guarantees the uniqueness of flat spherically symmetric foliations
for such spacetimes.
\end{quotation}


\section{Introduction}
Splitting a space into a sequence of subspaces, such that every
point in the space lies in one and only one of the subspaces, is
called a \emph{foliation}. The foliation of an $n$-dimensional
manifold, $M$, is a decomposition of $M$ into submanifolds, all
being of the same dimension, $p$. The submanifolds are the leaves of
the foliation. The co-dimension, $q$, of a foliation is defined as
$q=n-p$. A foliation of co-dimension one is called a foliation by
\emph{hypersurfaces}. The simplest and best understood cases of
foliation are when $p=q=1$, e.g. the two dimensional xy-plane,
$\mathbb{R^2}$, which can be foliated by the straight lines,
$y=mx+c$, with $c$ taken as the parameter and any fixed $m$. Notice
that a foliation of the of xy-plane by straight lines is not unique,
as different `fixed' values of $m$ will give different sequences of
foliating straight lines with a different slope.

In General Relativity (GR) one often requires to use a sequence of
spacelike or null hypersurfaces to foliate the spacetime. There
has been a lot of work to obtain foliations by hypersurfaces of
zero mean extrinsic curvature called `maximal slicing'
\cite{1}-\cite{4} and by hypersurfaces of constant mean extrinsic
curvature known as `CMC-slicing' \cite{4}-\cite{11}. There has
also been significant work on foliations by hypersurfaces of zero
intrinsic curvature called `flat foliations' \cite{12}-\cite{18}.
Existence of flat spacelike foliations for spherically symmetric
static spactimes (SSSS) is shown via the Hamiltonian equations of
general relativity in \cite{14} and using an initial value
approach in \cite{15}, \cite{16}. Complete foliations of the
Schwarzschild and Reissner Nordstr\"{o}m (RN) spacetimes by flat
spacelike hypersurfaces is also obtained using the fact that the
normals to such foliations are geodesics \cite{17}, \cite{18}.
Being indirect approaches, earlier procedures do not guarantee the
uniqueness of these foliations. As a flat foliation covers the
most interesting regions of spacetime describing realistic
gravitational collapse it is specially suited for studying Hawking
radiation from a fully quantum gravitational viewpoint. Husain and
Winkler \cite{19} have presented a flat slice Hamiltonian
formalism to have ``a standard model'' for studying black hole
physics. The non-uniqueness of flat foliations may raise the
question on the validity of the results if a different sequence of
flat slices is used in their model.

In this paper, in order to obtain flat spacelike hypersurfaces, we
use the direct approach (i.e. solve $R^{i}_{jkl} = 0$, where
$R^{i}_{jkl}$ are the components of the Riemann curvature tensor
for the hypersurfaces). Solution of the above system gives a
unique sequence of flat spherically symmetric spacelike
hypersurfaces admitted by SSSS, thus showing the uniqueness of
flat spherically symmetric foliations for such spacetimes. In the
following sections, after presenting solution of the equations
giving flat spherically symmetric hypersurfaces and some examples,
a conclusion is given.

\section{Flat Spherically Symmetric Hypersurfaces Admitted by Spherically
Symmetric Static Spacetimes}

The most general form of a spherically symmetric static spacetime
metric in the usual coordinates is
\begin{equation}
ds^{2}=e^{\upsilon \left( r\right) }dt^{2}-e^{\lambda \left(
r\right) }dr^{2}-r^{2}d\Omega ^{2},  \label{1}
\end{equation}
where
\begin{equation}
d\Omega ^{2}=d\theta ^{2}+\sin ^{2}\theta d\phi ^{2}.  \label{2}
\end{equation}
Now take an arbitrary hypersurface, $f(t,r,\theta ,\phi )=0$.
Considering spherical symmetry, taking $\theta $ and $\phi $
constant, this hypersurface in explicit form can be given as
\begin{equation}
t=F(r).  \label{3}
\end{equation}
The induced 3-metric (of the hypersurfaces) is then
\begin{equation}
ds_{3}^{2}=-\left( e^{\lambda \left( r\right) }-e^{\upsilon \left(
r\right) }F^{\prime 2}\right) dr^{2}-r^{2}d\Omega ^{2}.  \label{4}
\end{equation}
For the induced metric to be flat a \emph{necessary} but \emph{not
sufficient} condition, namely the Ricci scalar $=R=0$, implies
\begin{equation}
\frac{r\left( -\lambda ^{\prime }e^{\lambda }+\nu ^{\prime }e^{\nu
}F^{\prime 2}+2e^{\nu }F^{\prime }F^{\prime \prime }\right)
}{\left( e^{\lambda }-e^{\nu }F^{\prime 2}\right)
^{2}}+\frac{1-e^{\lambda }+e^{\nu }F^{\prime 2}}{e^{\lambda
}-e^{\nu }F^{\prime 2}}=0,  \label{5}
\end{equation}
where $^{\prime }$ represents the derivative with respect to $r$.
Using the substitution
\begin{equation}
g^{2}\left( r\right) =\frac{1}{e^{\lambda }-e^{\nu }F^{\prime 2}},
\label{6}
\end{equation}
Eq.$\left( \ref{5}\right) $ becomes
\begin{equation}
2rgg^{\prime }+g^{2}-1=0,  \label{7}
\end{equation}
and we have the general solution
\begin{equation}
g^{2}\left( r\right) =1-\frac{c}{r},
\end{equation}
where $c$ is an arbitrary constant with dimensions of length. The
induced metric now takes the form
\begin{equation}
ds_{3}^{2}=-\frac{dr^{2}}{1-\frac{c}{r}}-r^{2}d\Omega ^{2}.
\label{8}
\end{equation}
The above metric, Eq.$\left( \ref{8}\right)$, of the hypersurfaces
is flat, i.e. all the components of the Riemann curvature tensor
are zero (which is the necessary and sufficient condition for the
hypersurfaces to be flat), only if $c=0$ or in other words only if
\begin{equation}
g^{2}\left( r\right) =1.  \label{9}
\end{equation}
Then, from Eqs.$\left( \ref{3}\right) $ and $\left( \ref{6}\right)
,$ the flat spherically symmetric hypersurfaces are uniquely given
as
\begin{equation}
t=F\left( r\right) =\int e^{\frac{\lambda -\nu
}{2}}\sqrt{1-e^{-\lambda }}dr. \label{10}
\end{equation}
The mean extrinsic curvature, $K$, of these hypersurfaces is
\begin{equation}
K=e^{\left( \frac{\nu +\lambda }{2}\right) }\left( \frac{\nu
^{\prime }e^{\nu }}{2\sqrt{1-e^{\nu }}}-\frac{2\sqrt{1-e^{\nu
}}}{r}\right), \label{11}
\end{equation}
and the Hamiltonian constraint gives
\begin{equation}
R+K^{2}-K_{ab}K^{ab}=\frac{2(K^{2}-e^{\nu })}{r^{2}}-\frac{2\nu
^{\prime }e^{\nu }}{r}, \label{12}
\end{equation}
$($here for flat hypersurfaces $R=0)$.

\section{Some Examples}

For the exterior Schwarzschild spacetime, given by the metric in
Eq.$\left( \ref{1} \right)$ with $e^{\upsilon \left(
r\right)}=e^{-\lambda \left( r\right)}=1-2m/r$ where $m$ is the
mass, solution of Eq.$\left( \ref{10}\right)$ provides the unique
sequences of flat spherically symmetric spacelike hypersurfaces

\begin{equation}
t=F\left( r\right) =t_{c}-4m\sqrt{\frac{r}{2m}}-2m\ln \left|
\frac{\sqrt{ \frac{r}{2m}}-1}{\sqrt{\frac{r}{2m}}+1}\right| ,
\label{13}
\end{equation}
where $t_{c}$ is an integration constant which gives the
\emph{time} of the hypersurface i.e. the distinct values of
$t_{c}$ correspond to the distinct flat hypersurfaces. Notice that
the expression in Eq.$\left( \ref{12}\right) $ is same as the
Lema\^{i}tre coordinates (see, e.g., \cite{20}) for the
Schwarzschild geometry or the flat hypersurfaces obtained by using
the fact that these hypersurfaces are orthogonal to the unforced
geodesics in \cite{18}. The mean extrinsic curvature, $K$, of
these hypersurfaces is $3\sqrt{ \frac{m}{2r^{3}}}$. The
Hamiltonian constraint in this case gives
$R+K^{2}-K_{ab}K^{ab}=0$.

The exterior Reissner-Nordstrom spacetime is given by the metric
in Eq.$\left( \ref{1} \right) $ with $e^{\upsilon \left( r\right)
}=e^{-\lambda \left( r\right) }=1-2m/r+Q^{2}/r^{2}$ where $m$ and
$Q$ represent mass and charge respectively. In the case when $Q>m$
solution of Eq.$\left( \ref{10}\right)$ gives

\begin{equation}
t=F\left( r\right) =t_{c}-2E(r)-m\ln \left|
\frac{r-E(r)}{r+E(r)}\right| -
\frac{2m^{2}-Q^{2}}{\sqrt{Q^{2}-m^{2}}}\left[ \tan ^{-1}\left(
\frac{E(r)-m}{ \sqrt{Q^{2}-m^{2}}}\right) +\tan ^{-1}\left(
\frac{E(r)+m}{\sqrt{Q^{2}-m^{2}} }\right) \right] ,  \label{14}
\end{equation}
where
\begin{equation}
E(r)=\sqrt{2mr-Q^{2},}  \label{15}
\end{equation}
and $t_{c}$ is the constant of integration. For $Q<m$ we have
\begin{equation}
t=F\left( r\right) =t_{c}-2E(r)-m\ln \left|
\frac{r-E(r)}{r+E(r)}\right| -
\frac{2m^{2}-Q^{2}}{\sqrt{m^{2}-Q^{2}}}\ln \left[
\frac{mr-E(r)\sqrt{
m^{2}-Q^{2}}-Q^{2}}{mr+E(r)\sqrt{m^{2}-Q^{2}}-Q^{2}}\right] ,
\label{16}
\end{equation}
and for the extreme case i.e. $Q=m$ we have
\[
t=F\left( r\right) =t_{c}-2E(r)+\frac{mE(r)}{r-m}+4m\tanh
^{-1}\left[ \frac{ E(r)}{m}\right] .
\]

The mean extrinsic curvature, $K$, of the flat  hypersurfaces in
all cases of the RN spacetime is
$\frac{3mr-Q^{2}}{r^{2}\sqrt{2mr-Q^{2}}}$ and the Hamiltonian
constraint gives $R+K^{2}-K_{ab}K^{ab}=\frac{2Q^{2}}{r^{4}}$.

\section{Conclusion}

There has been lot of work on existence and construction of
foliation of SSSS by spacelike hypersurfaces of zero intrinsic
curvature. Perhaps, assuming the difficulty to solve the system of
differential equations, $R^{i}_{jkl} = 0$, in all earlier works
indirect approaches have been used. In this paper, in order to
obtain all possible sequences of flat spherically symmetric
hypersurfaces admitted by SSSS, we have solved this system of
differential equations. It is found that there exists a unique
sequence of flat spherically symmetric spacelike hypersurfaces
admitted by SSSS, guaranteeing the uniqueness of foliation by
these hypersurfaces for such spacetimes. To emphasize the point,
it is not just that the foliation of SSSS by flat spherically
symmetric spacelike hypersurfaces is unique, but that these
spacetimes admit a unique sequence of flat spherically symmetric
spacelike hypersurfaces which form a foliation. Notice that the
flat spherically symmetric spacelike hypersurfaces can also be
obtained simply by changing the sign in Eq.$\left( \ref{3}\right)$
and in expressions for extrinsic curvatures. This corresponds to
the hypersurfaces orthogonal to the incoming instead of outgoing
geodesics \cite{18}.

In this paper we have studied flat slices of static, spherically
symmetric spacetimes, where the slices themselves are also assumed
to have spherical symmetry. Dropping the latter restriction would
result in a study of a system of partial differential equations
for the height function, as opposed to the ordinary differential
equation, Eq.$\left( \ref{5}\right)$, of the slice. This is
outside the scope we have set ourselves here.

\section{Acknowledgments}

One of the authors AAS highly appreciates role of the Higher
Education Commission of Pakistan and is thankful for the financial
support during the post doctoral research in Sweden during which
this work was initiated. Authors are also grateful to the two
referees for their comments which have improved quality of the
paper.

\end{document}